# Thermosensitive Cu$_2$O-PNIPAM core-shell nanoreactors with tunable photocatalytic activity

He Jia,[1] Rafael Roa,[1] Stefano Angioletti-Uberti,[2,1] Katja Henzler,[3] Andreas Ott,[1] Xianzhong Lin,[4] Jannik Möser,[5] Zdravko Kochovski,[1] Alexander Schnegg,[5] Joachim Dzubiella,[1,6] Matthias Ballauff,[1,6] Yan Lu[1]*

[1]*Soft Matter and Functional Materials, Helmholtz-Zentrum Berlin für Materialien und Energie, Hahn-Meitner-Platz 1, Berlin (Germany)*
[2]*International Research Centre for Soft Matter, Beijing University of Chemical Technology, Beisanhuan East Road 19, 100099 Beijing (PR China)*
[3]*Paul Scherrer Institut, 5232 Villigen PSI (Switzerland)*
[4]*Heterogeneous Materialien und Energie, Helmholtz-Zentrum Berlin für Materialien und Energie, Hahn-Meitner-Platz 1, Berlin (Germany)*
[5]*Institute for Nanospectroscopy Helmholtz-Zentrum Berlin für Materialien und Energie, Kekulestr 5, Berlin (Germany)*
[6]*Institut für Physik, Humboldt-Universität zu Berlin, Newtonstr. 15, Berlin (Germany)*

We report a facile and novel method for the fabrication of Cu$_2$O@PNIPAM core-shell nanoreactors using Cu$_2$O nanocubes as the core. The PNIPAM shell not only effectively protects the Cu$_2$O nanocubes from oxidation, but also improves the colloidal stability of the system. The Cu$_2$O@PNIPAM core-shell microgels can work efficiently as photocatalyst for the decomposition of methyl orange under visible light. A significant enhancement in the catalytic activity has been observed for the core-shell microgels compared with the pure Cu$_2$O nanocubes. Most importantly, the photocatalytic activity of the Cu$_2$O nanocubes can be further tuned by the thermosensitive PNIPAM shell, as rationalized by our recent theory.

## INTRODUCTION

Cu$_2$O is a well-known p-type semiconductor with direct band gap of 2.17 eV. It has a great potential for a wide range of applications, e.g. in solar energy conversion, lithium-ion batteries, gas sensors, photocatalytic degradation of dye molecules, propylene oxidation and photoactivated water splitting. The properties of the Cu$_2$O nanoparticles are strongly dependent on their shape. Hence, there is a growing interest in the synthesis of Cu$_2$O nanostructures with defined shape.[1–6] Thus, Cu$_2$O nanocubes, octahedral, nanocages, spheres, nanowires and other highly symmetrical structures have already been reported.[7,8]

A main drawback for further applications of Cu$_2$O nanoparticles is that Cu$_2$O is easily oxidized in water and the nanostructure of Cu$_2$O can be destroyed depending on external conditions such as pH or visible light. For this reason, a simple and effective method providing protection of Cu$_2$O-based nanostructures from oxidation is highly desirable. Parecchino et al. successfully improved the chemical stability of a Cu$_2$O layer in water through atomic layer deposition of multiple protective layers of Al-doped zinc and titanium oxide.[9,10] Wang's group reported that both CuO and carbon can be used to protect Cu$_2$O films and nanofibers.[11,12] Notably, the aforementioned protection strategies have all been applied to extended one- and two-dimensional phases. However, little has been reported in the literature regarding the effective protection of Cu$_2$O nanoparticles. In this regard, Yang et al.[13] and Su et al.[14] have successfully synthesized Cu$_2$O@SiO$_2$ core-shell nanoparticles, but unfortunately the SiO$_2$ shell makes them aggregate more easily, preventing further study on their surface properties.

Recently, shells of poly(*N*-isopropylacrylamide) (PNIPAM) core-shell microgels have been used to modify inorganic nanoparticles.[15,16] In this way, the nanoparticles encapsulated inside PNIPAM shells can be prevented from aggregation in aqueous solution.[17] For example, Zhao[18] and co-workers reported the fabrication of gold nanoparticles with a thin PNIPAM shell, proposed as a drug delivery system. Of great relevance for catalytic applications is also the fact that the catalytic properties of the embedded nanoparticles can be tuned by the swelling and deswelling of the PNIPAM microgels.[19–22] In this regard, Liz-Marzán et al.[23] developed different kinds of core-shell hybrid systems via growth of PNIPAM gels on the surface of metal nanoparticles. Recently, some of us presented a theory for the diffusion- and solvation-controlled contribution to the reaction rate of such a "nanoreactor".[24] There, it was demonstrated that the thermosensitive shell can be used to enhance or reduce the local concentration and permeability of a given reactant and thus increase or decrease the total catalytic activity of the embedded nanoparticle, respectively. Thus, core-shell systems consisting of a catalytically active nanoparticle and a polymeric shell present a novel type of nanoscopic catalyst with tunable properties.

To the best of our knowledge, until now no work has been reported on colloidal stable Cu$_2$O nanoparticles modified with PNIPAM shells. Such functional hybrid nanoparticles will not only improve the stability of sensitive semiconductor nanoparticles, but can also be applied as "nanoreactors" with stimuli-responsibility. Here we present for the first time the synthesis and characterization of Cu$_2$O@PNIPAM core-shell hybrid nanoparticles using cubic-shaped Cu$_2$O nanoparticles as core. Scheme 1 shows the procedure of the main synthesis step for this system, i.e. coating of the nanoparticle with the PNIPAM shell: Without modification with a SiO$_2$ or polystyrene interlayer [25-27], a single Cu$_2$O nanocubes is encapsulated in a thermosensitive PNIPAM shell which prevents aggregation. The synthesis proceeds in two steps: First, the surface of the Cu$_2$O-cubes is modified by an interlayer of poly(diallyldimethylammonium chloride) (PDDA) and sodium styrene sulfonic acid (NaSS). In a second step, the PNIPAM-shell is attached to the cubes by a precipitation polymerization. We demonstrate that the photocatalytic activity of the Cu$_2$O nanocubes is significantly enhanced by the PNIPAM shell, and can be further tuned by temperature via the thermosensitive shell, as suggested by theory.[24] The present work opens a new way for the surface modification of Cu$_2$O nanocubes, which will have a great potential for applications of Cu$_2$O nanoparticles. In addition, such core-shell "nanoreactor" system is essential to understand the effect of PNIPAM shell on properties of metal or metal oxide nanomaterials.

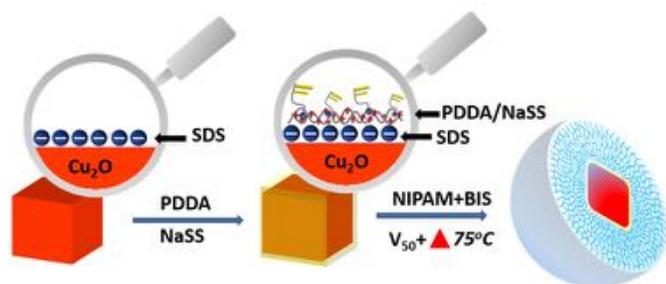

**Scheme 1.** Schematic illustration of the procedure used to coat PNIPAM on the surface of Cu$_2$O nanocubes. With sodium dodecyl sulfate (SDS) as the surfactant, the surfaces of Cu$_2$O nanocubes are negatively charged. By charge interaction, NaSS was modified on the surface of Cu$_2$O nanocubes using PDDA as a medium. Under high temperature, polymerization is initiated by the positive initiator V50 and PNIPAM-shell is coated around Cu$_2$O nanocubes.

## Experimental Section

### MATERIALS AND METHODS

**Materials:** Copper chloride (CuCl$_2$), sodium dodecyl sulfate (SDS), sodium hydroxide (NaOH), sodium ascorbate, poly(diallyldimethylammonium chloride) (20 wt.% in H$_2$O) (PDDA), 4-styrenesulfonic acid sodium salt hydrate (NaSS), methyl orange (MO), 5,5-dimethyl-pyrrdine N-oxide (DMPO), N-isopropylacrylamide (NIPAM) and N,N'-methylene-bis-acrylamide (BIS) were supplied by Aldrich. 2,2'-azobis(2-methylpropionamidine dihydrochloride) (V50) was supplied by Fluka. All of the reactants were used without further purification. Water was purified by a Milli-Q system.

**Synthesis of Cu2O nanocubes:** Cu$_2$O nanocubes with size of 259±19 nm were synthesized by seed-mediated reaction modified with the method reported by Michael H. Huang's group (the size distribution has been shown in Figure s1 in the Supporting Information).[28] Briefly, at first, a volume of 10 mL aqueous solution containing $10^{-3}$ M CuCl$_2$ and $3.3 \times 10^{-2}$ M sodium dodecyl sulfate (SDS) was prepared. Then 250 µL of 0.2 M sodium ascorbate and 500 µL of 1 M NaOH solution was added sequentially with shaking for 5 s to prepare the seeds solution. 1 mL of this seeds solution was transferred to 9 mL solution containing $10^{-3}$ M CuCl$_2$ and $3.3 \times 10^{-2}$ M SDS with shaking for 10 s as the seeds solution for the next step. The same process was repeated for three times. In the last step, the volume of the reaction solution was scaled up to 81 mL. 9 mL seeds solution from the previous step was transferred into it. After shaking for 15 s, 2.25 mL of 0.2 M sodium ascorbate and 4.5 mL of 1M NaOH was added separately with shaking for 8 s. After standing at room temperature in a dark place for 2 h, the color of the solution became orange due to the formation of Cu$_2$O nanocubes. Then, the Cu$_2$O nanocubes were washed by centrifugation with a speed of 3500 rpm in water for 20 min and dispersed into 10 mL H$_2$O.

**Synthesis of Cu2O@PNIPAM core-shell nanoparticles:** In the first step, Cu$_2$O nanocubes were modified with PDDA and NaSS as follows. 2.6 g PDDA was diluted with 27.5 mL H$_2$O and then 10 mL NaSS (0.024 M) solution was added with the rate of 20 mL/h. After stirring for another 2 hours at a speed of 500 rpm, 10 mL Cu$_2$O nanocubes solution was added into the mixed solution slowly. The excess PDDA and NaSS were removed by centrifugation (3500 rpm, 15 min) and the modified Cu$_2$O nanocubes were dispersed into 5 mL H$_2$O. The Cu$_2$O@PNIPAM core-shell nanoparticle was prepared by precipitation polymerization. Under continuous vigorous stirring and nitrogen atmosphere, the solution containing the modified Cu$_2$O nanocubes was heated to 75 °C. Thereafter, 1 mL V50 (0.018M) solution was added drop by drop as the initiator. The polymerization was started immediately with the addition of 29 mg NIPAM and 5.1 mg BIS (dissolved in 1 mL H$_2$O). The orange solution became turbid after 10 min and the reaction was run for 2 h. The composite particles were then purified by centrifugation and redispersion in water several times.

### CHARACTERIZATION

The hydrodynamic radius of the samples as a function of temperature was conducted by Zetasizer (Malvern Zetasizer Nano ZS ZEN 3500). The UV-vis spectra were measured by a Lambda 650 spectrometer supplied by Perkin-Elmer or Agilent 8453 with a temperature controlled sample holder with an accuracy of ±0.1 °C. Transmission electron microscope (TEM) images were done with JEOL JEM-2100 at 200kV. XRD measurements were performed in a Bruker D8 diffractometer in the locked coupled mode (scanning angle from 10° to 90°) with Cu Kα1 radiation, the incident wavelength is 1.5406 Å. For the accomplished measurements the acceleration voltage was set to 40 kV and the filament current to 40 mA. The amount of $Cu_2O$ in the $Cu_2O$@PNIPAM core-shell nanoparticles was determined by thermogravimetric analysis (TGA) using a Netsch STA 409PC LUXX. Fifteen milligrams of dried sample were heated to 800 °C under a constant argon flow (30 mL/min) with a heating rate of 10 K/min. Scanning electron microscope (SEM) measurements were done in a SEM LEO GEMINI 1530. The size and size distribution of $Cu_2O$ nanoparticles were measured using Image J software based on their TEM images. At least 100 units were counted.

**Photocatalytic measurements:** For testing the photocatalytic activity, 10 mL $Cu_2O$@PNIPAM nanoparticles (0.10 wt. %) were dispersed into 90 mL of an aqueous solution containing 15.6 mg/L methyl orange. The samples were first stirred in the dark for 30 min in order to ensure the adsorption of MO into the $Cu_2O$@PNIPAM core-shell nanoparticles. A 500 W xenon lamp was used as the light source, which was placed 20 cm away from the samples. UV-vis absorption spectra of the samples were taken every 30 min by removing the cap to withdraw the solution. The temperature of the reaction was controlled by a water bath with an accuracy of ± 0.2°C. The reaction rate $k$ can be defined through normalization of $k_{app}$ to the total surface of the $Cu_2O$ nanocubes in the system. TGA (see Figure s2) and TEM results have been used to obtain the amount and size of $Cu_2O$ nanocubes in the core-shell nanoparticles. For the calculation of the surface area ($S$) of the $Cu_2O$ nanocubes in the core-shell nanoparticles, the density of $Cu_2O$ (6.00 g $cm^{-3}$) was used.

**Cryogenic transmission electron microscopy:** Cryo-TEM specimens were vitrified by plunging the samples into liquid ethane using an automated plunge freezer (Vitrobot Mark IV, FEI). The lacey carbon copper grids (200 mesh, Science Services) have been pretreated by 10 seconds of glow discharge and equilibrated for 5 minutes at 15 °C or 50 °C inside the plunge freezer. Approximately 5 μl of a pre-temperatured 0.025 wt. % solution were given on the TEM grids and equilibrated at the adjusted temperature for 2 minutes in a water-saturated atmosphere. After blotting the liquid, the specimen were vitrified, inserted into a pre-cooled Gatan 914 sample holder and transferred into a JEOL JEM-2100, operating at 200kV.

**Near edge X-ray absorption fine structure – transmission X-ray microscopy (NEXAFS-TXM):** Sample preparation for NEXAFS-TXM: [29] The carbon coated copper grids have been pretreated by 10 s of glow discharge. Approximately 5 μl of a 0.1 wt% dispersion of the particles was deposited on a TEM copper grid with a carbon support film (200 meshes, Science Services, Munich, Germany). The grids were dried at room temperature. The NEXAFS-TXM spectra were recorded on the O-$K$-edge and the Cu-$L_{2,3}$-edge with the HZB-TXM which is installed at the undulator beamline U41-FSGM at the electron storage ring BESSY II, Berlin, Germany. It provides a high spatial resolution close to 10 nm (half-pitch) and a spectral resolution up to E/ΔE ≈ $10^4$. Typical spectra are presented for each set of measurements. The TXM allows measurements to be taken at room or liquid nitrogen temperature in a vacuum of 1.3x$10^{-9}$ bar. The spectra were recorded at room temperature in transmission mode by taking a sequence of images over a range of photon energies covering the investigated absorption edges with a calculated E/ΔE > 5800 for the Cu-$L_{2,3}$-edge and E/ΔE > 12000 for the O-$K$-edge. Note that the exit slit of the monochromator was set to 9 μm for the Cu-L2,3-edge and 7 μm for the O-$K$-edge resulting in the given calculated monochromaticy values. The exposure time for one image with 1340×1300 pixels was 40 s for the Cu-$L_{2,3}$-edge and 4 s for the O-$K$-edge to achieve a sufficient signal to noise ratio in the images. Taking an image stack with up to 226 images at different energies needs inherently about 45 to 120 min because of all necessary movements, exposure time, and camera read out time and image storage. The NEXAFS spectra were normalized since the photon flux varies as a function of photon energy (hv) and time in the object field (x, y). The normalization was performed by dividing the intensity I(x, y, hv) recorded on a single nanostructure by the intensity $I_0$ (x+Δx, y+Δy, hv) recorded in its sample free proximity at position (x+Δx, y+Δy). Both I(x, y, hv) and $I_0$(x, y, hv) were recorded within the same image stack since bare regions in the vicinity of the nanostructures permit the measurement of $I_0$.

**Electron spin resonance (ESR):** Continuous wave ESR (cwESR) spectra were obtained on a Bruker ESP 300 spectrometer with a Bruker ER-4122 super high quality factor (SHQ) resonator at room temperature. During the ESR measurements samples immersed in the ESR resonator were illuminated through a 250 W cold halogen lamp (Schott KL 2500 LCD). For spin trapping 5,5-dimethyl-pyrroline N-oxide (DMPO) was used. A 50 μL aqueous solution of 50 mg/mL DMPO was mixed with 50 μL of $Cu_2O$ nanoparticle solution with a concentration of 0.20 mg/mL for the $Cu_2O$@PNIPAM core-shell microgels and 0.15 mg/mL for the pure $Cu_2O$ nanocubes, respectively, in order to ensure an equal amount of $Cu_2O$ in both samples. 20 μL of the mixed sample solution was filled into a Q-band ESR sample tube (inner diameter 1 mm). The resonator was critically coupled yielding quality factors of 4000-5000. Magnetic field modulation for phase-sensitive detection by means of a lock-in amplifier was employed at a frequency of 100 kHz and a peak-to-peak amplitude of 1 G. An incident microwave power of 2 mW was used for all measurements. Spectra were normalized by resonator quality and sample volume.

## RESULTS AND DISCUSSION

Prior to the coating step, $Cu_2O$ nanocubes were initially prepared by the seed-mediated method using sodium dodecyl sulfate (SDS) as the capping surfactant.[28] As shown in Figure 1a, $Cu_2O$ nanocubes with size of 259±19 nm were synthesized. As the interlayer, PDDA with NaSS was adsorbed due to the charge attraction. The ratio of positive and negative charges contained in PDDA and NaSS is 1:0.074. Thus, after mixing with NaSS, there are still large amounts of positive charges in PDDA chains. PDDA/NaSS is firmly attached to the negatively-charged surface of the SDS stabilized cubes as can be seen from the surface zeta potential which changed from -21.2 mV to 53 mV.

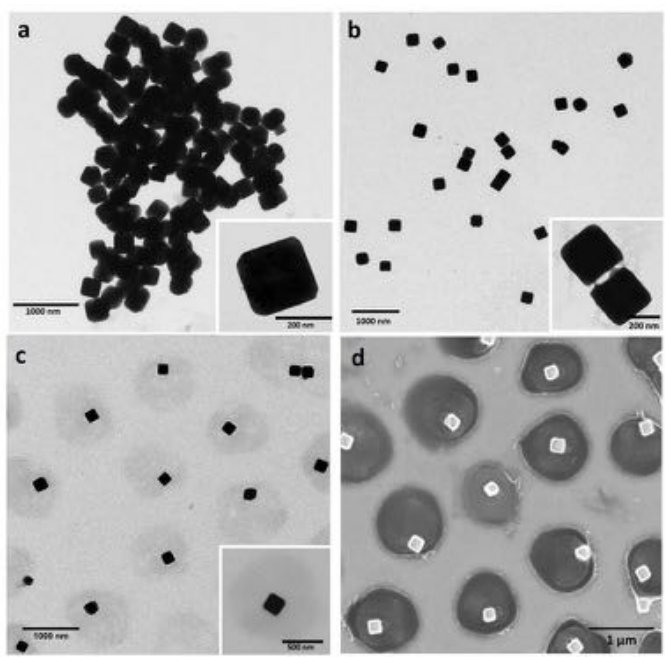

**Figure 1.** TEM images of (a) $Cu_2O$ nanocubes, (b) $Cu_2O$@PDDA-NaSS, (c) $Cu_2O$@PNIPAM core-shell nanoparticles; (d) SEM image of $Cu_2O$@PNIPAM core-shell nanoparticles.

The double bonds supplied by the NaSS attached to the surface help the chemical bonding of the PNIPAM shell on the $Cu_2O$ core surface via a precipitation polymerization. Without this surface modification, most of the $Cu_2O$ nanocubes aggregate quickly.

As shown in Figure 1b, well-dispersed $Cu_2O$ nanocubes were obtained after PDDA/NaSS coating, which was due to the large number of positive charges provided by PDDA. From the inset TEM image in Figure 1b, a very thin PDDA/NaSS layer can be observed clearly. The successful coating of PNIPAM shell on the $Cu_2O$ nanocubes surface was first confirmed by TEM and SEM images as shown in Figure 1c and 1d. The PNIPAM shell was uniformly wrapped around the surface of $Cu_2O$ nanocubes and all of the $Cu_2O$ nanoparticles retained their cubic shapes. Since PNIPAM promotes strong hydrophilic repulsion below the lower critical solution temperature (LCST)[30-35], the $Cu_2O$@PNIPAM core-shell nanoparticles are well separated from each other, as can be seen clearly from the TEM image of $Cu_2O$@PNIPAM in Figure 2a. Figure 2b shows a cryo-TEM image of the particles prepared at room temperature. The thickness of the PNIPAM shell was about 360 nm, which agreed well with the thickness of the PNIPAM shell determined by DLS at room temperature in water (marked as a dashed line in Figure 2b).

DLS measurements of $Cu_2O$@PNIPAM core-shell nanoparticles shown in Figure 2c proved the thermosensitivity of PNIPAM shell, as expected. A well-defined volume phase transition is observed around 32 °C for the $Cu_2O$@PNIPAM core-shell system. Figure 2d and Figure 2e show the cryo-TEM images of the core-shell nanoparticles at 15 °C and 50 °C, which are in perfect agreement with the DLS data (see point d and point e on the DLS curve Fig. 2c). Below the LCST, the PNIPAM shell is fully swollen in the water solution (see Figure 2d). When the temperature is increased to 50°C, the water in the PNIPAM network is extruded, leading to the shrinkage of the PNIPAM shell (Figure 2e).

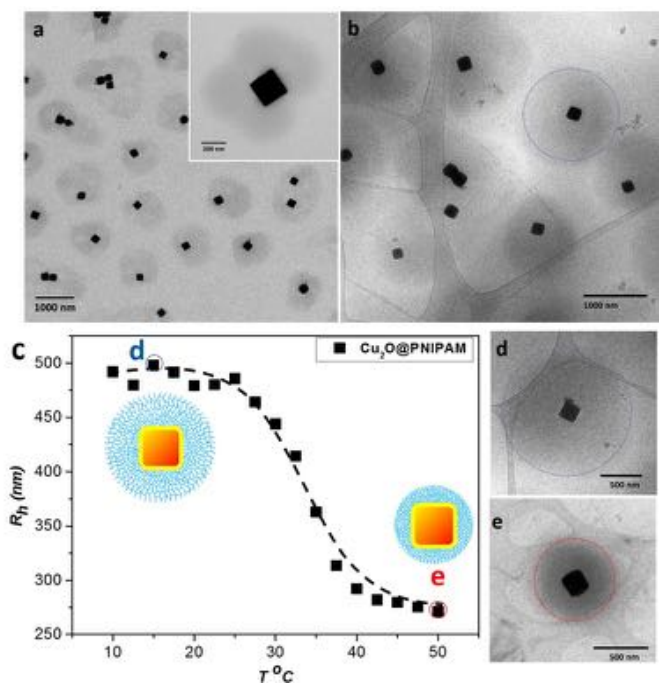

**Figure 2.** (a) Overview TEM image of $Cu_2O$@PNIPAM core-shell nanoparticles, (b) Cryo-TEM image of $Cu_2O$@PNIPAM core-shell nanoparticles (Dashed circle indicate the size of the PNIPAM shell in swollen state at room temperature determined by DLS), (c) Hydrodynamic radius of $Cu_2O$@PNIPAM core-shell nanoparticles as a function of temperature in aqueous solution, (d,e) Cryo-TEM images of $Cu_2O$@PNIPAM core-shell nanoparticles in swollen state at 15 °C and in shrunken state at 50 °C, respectively.

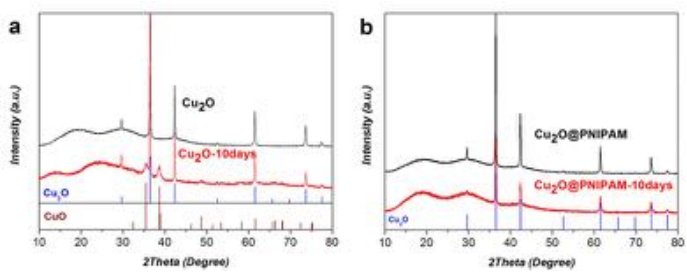

**Figure 3.** XRD patterns of (a) $Cu_2O$ nanocubes without coating PNIPAM: kept in water at room temperature for 10 days (red) and fresh prepared (black). (b) $Cu_2O$@PNIPAM core-shell nanoparticles: kept in water at room temperature for 10 days (red) and fresh prepared (black). As reference, standard XRD patterns of CuO (JCPDS: No.45-0937) and $Cu_2O$ (JCPDS: No.65-3288) are shown in the Figures.

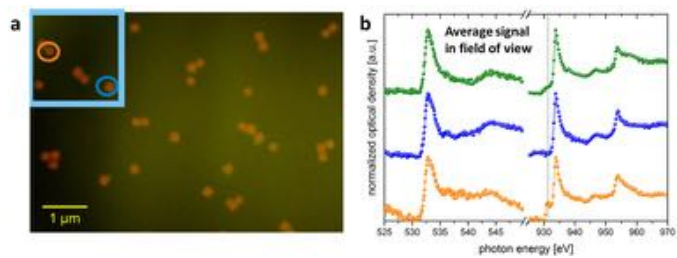

**Figure 4.** (a) TXM micrograph of the $Cu_2O$-nanocubes@PNIPAM at two different photon energies: The red channel depicts the nanocubes with $Cu_2O$, and the green channel refers to CuO. (b) NEXAFS-spectra of the average signals over all $Cu_2O$@PNIPAM core-shell nanoparticles in field of view (green line) and the marked particles in the inset of micrograph on the left hand side at the O-$K$-edge and the Cu-$L_{2,3}$-edge (blue and orange lines).

The XRD patterns of $Cu_2O$ nanocubes without PNIPAM shell are presented in Figure 3a. The peaks at $2\theta$= 29.63°, 36.50°, 42.40°, 52.58°, 61.52°, 73.70° and 77.57° correspond to the Bragg reflections of $Cu_2O$ nanocrystals indicating the production of fresh $Cu_2O$. After storing in water in a dark place for 10 days, new peaks at $2\theta$= 35.5°, 38.34°, 38.66° and 48.8° corresponding to the Bragg reflections of CuO are observed. This indicates the cubic structure of the $Cu_2O$ has been destroyed due to the formation of CuO.

Meanwhile, the color of the solution changed from orange to dark green as shown in Figure s4a in the supporting information. Compared with the freshly prepared $Cu_2O$ nanocubes, PNIPAM-modified nanocubes also preserve all of the characteristic $Cu_2O$ peaks, and no other peaks were found, indicating that none of the $Cu_2O$ in these particles was oxidized during the process of coating with PNIPAM (as shown in Figure 3b). In addition, after storing for 10 days under the same condition used with pure $Cu_2O$ nanocubes, no $Cu_2O$@PNIPAM core-shell nanoparticles were oxidized to CuO, as proved by the XRD patterns in Figure 3b. Moreover, all of the $Cu_2O$ maintained the cubic structure and no aggregation was observed in the system (see Figure s4b). Obviously, the PNIPAM-shell is capable of preventing the oxidation of $Cu_2O$ in an efficient manner. A further proof of the protection by the PNIPAM shell can be given by NEXAFS measurement for the $Cu_2O$@PNIPAM core-shell nanoparticles, which have been kept in water at room temperature for 100 days. Near edge X-ray absorption fine structure spectroscopy (NEXAFS) in combination with transmission X-ray microscopy (TXM) has been used to test the samples for the existence of $Cu^{2+}$ at crystal defect sites or amorphous CuO. The NEXAFS spectroscopy provides chemical sensitivity and the TXM gives the possibility to analyze single particles.

Figure 4a shows a NEXAFS-TXM micrograph in false color representation. The red channel is sensitive to $Cu_2O$ whereas the green channel is sensitive to CuO. The contrast of the PNIPAM shell is too low to be resolved. Figure 4a demonstrates that almost all particles consist of copper-(I)-oxide. Only very few small green points on the surface of the $Cu_2O$ nanocubes can be found in Figure 4a, which leads to the small shoulder around 930.8 eV in the spectrum of orange marked particle as shown in Figure 4b. A similar shoulder is also found in the spectra of freshly synthesized $Cu_2O$ nanoparticles (Figure s5).

Thus, this signal is most probably related to side products from the synthesis of the $Cu_2O$ nanocubes, which could not be removed during the purification step or from the preparation process for the TXM measurement. In addition, for both $Cu_2O$@PNIPAM core-shell nanoparticles and the bare $Cu_2O$ nanocubes, this shoulder became much smaller in the spectrum of the average signals over all particles in field of view confirming the extremely low proportion of $Cu^{2+}$ in the systems. The NEXAFS-TXM results indicate that the PNIPAM shell can protect effectively the $Cu_2O$ nanocubes from oxidation for at least months.

Figure s6 shows TEM images of individual $Cu_2O$ nanocubes before and after PNIPAM coating and their corresponding selected area electron diffraction (SAED) patterns. The TEM images are provided to show the exact orientations of the $Cu_2O$ nanocubes for the SAED patterns. The circular field of view results from the SA aperture that was inserted to show the actual regions of the sample contributing to the diffraction patterns. These were indexed by comparison with simulated {100} $Cu_2O$ single crystal diffraction patterns using CrystalMaker and SingleCrystal (CrystalMaker Software Ltd, Oxfordshire, UK). The SAED patterns shown in Figure s6 a2 and b2 directly demonstrate that after the modification with PNIPAM, the nanocubes are still $Cu_2O$ without oxidation, since the spacings are not consistent with the monoclinic

structure of CuO. The PNIPAM coating does not create additional diffractions spot due to its amorphous structure. The photocatalytic activity of Cu$_2$O@PNIPAM core-shell nanoparticles has been determined by monitoring the photodegradation of methyl orange (MO) under visible light. Here, the decay of the strong absorption at 464 nm is measured as a function of time (Figure 5a; see also Figure s7). The pure Cu$_2$O nanocubes are not photocatalytically active at 15 °C as shown in Figure 5b.[8]

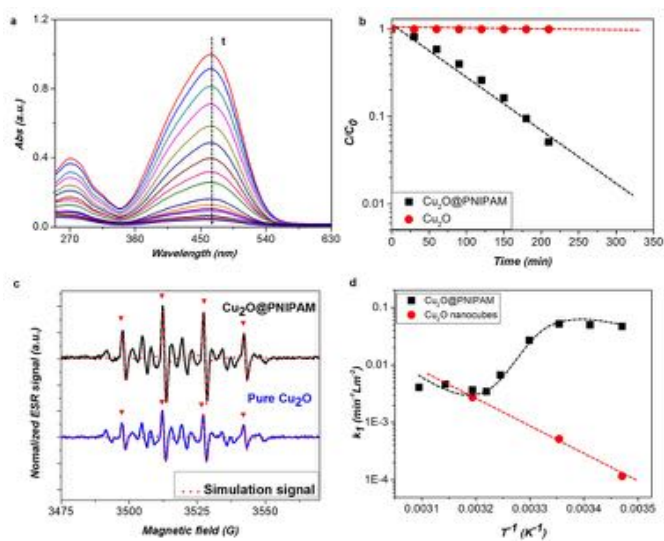

**Figure 5.** (a)UV-vis absorption spectra of MO as a function of irradiation time using Cu$_2$O@PNIPAM core-shell nanoparticles as the photocatalyst. (b) Kinetic analysis of MO reduced by Cu$_2$O@PNIPAM core-shell nanoparticles and pure Cu$_2$O nanocubes at room temperature. (c) ESR spectra of aqueous dispersions of Cu$_2$O@PNIPAM core-shell nanoparticles (upper black trace) and pure Cu$_2$O nanocubes (lower blue trace) with DMPO spin traps added. Triangles indicate the four ESR lines characteristic DMPO-•OH radicals. The overlaid traces are simulations of the DMPO-•OH signal using the MATLAB library EasySpin. (d) The reaction rate $k_1$ (rate constant $k_{app}$ normalized to the surface area of Cu$_2$O nanocubes[24]) at different temperatures for the bare Cu$_2$O nanocubes (circles) and the Cu$_2$O@PNIPAM core-shell nanoreactors system (squares).

To compare the photocatalytic activity of different systems directly, the reaction rate $k$, i.e. the experimentally measured reaction rate normalized to the total surface of the Cu$_2$O nanocubes, has been applied. As we previously showed, this normalized quantity can be directly linked to the catalytic properties of a single particle.[24] It is found that the reaction rate is extremely small for the pure Cu$_2$O nanocubes at 15 °C, as expected ($k_{Cu2O-15°C}$ = 1.16×10$^{-4}$ Lmin$^{-1}$m$^{-2}$,). Cu$_2$O nanocubes contain mostly {100} facets which have 100% saturated oxygen bonds. These {100} facets are neutral and there is no strong driving force for the adsorption of the negatively charged dye molecules onto the surface of the cubic Cu$_2$O nanocrystals.[7,36] Thus, the low reaction rate observed here is in accordance with similar observations in the literature.[7,37,38]

Instead, after modification by a PNIPAM shell, the photocatalytic activity of Cu$_2$O nanocubes was significantly enhanced ($k_{Cu2O@P-15°C}$=4.67×10$^{-2}$ Lmin$^{-1}$m$^{-2}$). Figure 5b displays the temporal decay of MO as the function of time for the Cu$_2$O@PNIPAM core-shell nanoparticles and the pure Cu$_2$O nanocubes at 15 °C. After irradiation for 4 h, the remaining amount of MO was ca. 3% of the initial amount in the presence of the Cu$_2$O@PNIPAM core-shell nanoparticles, compared to 99.4% in the case of pure Cu$_2$O nanoparticles.

The significantly enhanced photocatalytic efficiency of Cu$_2$O@PNIPAM core-shell systems over the bare nanoparticles is not due to a single factor, but rather come from a combination of many of them. Firstly, the colloidal stability of Cu$_2$O@PNIPAM is much higher than that of pure Cu$_2$O nanocubes and no aggregation is seen for the core-shell Cu$_2$O@PNIPAM particles during the photocatalytic reaction, thus leading to a larger effective catalytic area for the reaction to occur. Furthermore, adsorption data of reactants to the nanoreactors (see Figure s8) also indicate a high binding affinity of the hydrogel shell, with average reactant concentrations two orders of magnitude higher in the shell than in the bulk. Zeta potential measurements for the core-shell systems at 15 °C, showing a decrease of +13.7 mV to -5.3 mV upon adding the negatively charged MO to the suspension, support the picture of MO enrichment in the shell. This result strongly suggests that also a higher local concentration of MO adjacent to the nanoparticle is achieved. As we find that the reaction is fully surface-controlled (i.e., the reactant diffusive transport is much faster than the surface reaction, see supporting information), MO enrichment at the nanocube surface would lead proportionally to a higher surface rate.[24] However, details on such a rate enhancement depend on the exact spatial partitioning of the MO molecules in the nanoreactor that is experimentally not easily accessible. Another reason for the enhanced rate is related to the stronger absorption of light and narrower band gap in case of the core-shell particles (see Figure s9).[39-41] This in turn leads to a higher concentration of hydroxyl radicals close to the nanocubes surface, which are the active species in the process of photodegradation of MO.[42,43] As our reaction is fully surface-controlled, the total rate is directly proportional to the surface concentration of the hydroxyl radicals, and thus increases.

In detail, ESR spin trapping with DMPO was employed[44] to detect the concentration of hydroxyl radicals in illuminated solutions for pure Cu$_2$O nanocubes as well as for Cu$_2$O@PNIPAM core-shell nanoparticles. Both traces in Figure 5d contain a large variety of ESR resonances. Among them 4 peaks with relative intensities of 1:2:2:1 at g=2.0054 with a hyperfine splitting of $a_N$=$a_{Hβ}$=15.0 G can be clearly assigned to the DMPO-•OH adducts. The assignment was confirmed by a simulation of the DMPO-•OH signal by means of the MATLAB simulation toolbox EasySpin[45] (see Figure 5c). These lines are present in both solutions upon illumination. For the Cu$_2$O@PNIPAM core-shell nanoparticles the DMPO-•OH signal is at least two times stronger as compared to pure Cu$_2$O nanocubes. This indicates that much more •OH have been effectively generated by Cu$_2$O@PNIPAM core-shell nanoparticles than by pure Cu$_2$O nanocubes.

The influence of temperature on the photocatalytic activity has been studied as well. As shown in Figure 5d, compared to the rate constant of bare Cu$_2$O nanocubes at different temperatures, the reaction rate of Cu$_2$O@PNIPAM core-shell nanoreactors does not follow a simple Arrhenius law with constant activation energy. Below the LCST, the reaction rate of Cu$_2$O@PNIPAM core-shell system increased with rising temperature and showed a local maximum at around 25 °C. If the temperature was increased close to the LCST, a dramatic decrease of the reaction rate was observed. Then a slight increase of the reaction rate took place with further increasing temperature.

The reasons for the temperature induced change of the reaction rate can be qualitatively rationalized by the following arguments. First, the overall physicochemical properties of the Cu$_2$O@PNIPAM core-shell nanoreactors have a strong temperature dependence. Above LCST the gel is in a collapsed, hydrophobic state, which strongly suggests a local re-partitioning of MO within the PNIPAM-shell. In fact, MO is known to situate itself at the interface between water and oil in a surfactant-like style,[46] meaning that MO could be mostly located in a thin region constituting the hydrogel/solvent interface (see Figure 6). This view of high adsorption of MO to the hydrogel surface is also supported by the measured zeta potential of the nanoreactors at 40 °C that substantially decreases from a positive +22 mV to a negative -14.3 mV after addition of MO. The re-partitioning and enrichment of MO at the hydrogel/solvent interface, in turn, would lead to a reduced MO concentration adjacent to the nanoparticle. According to our theory[24] this depletion of reactants should effectively reduce the rate. Secondly, from the UV-vis spectra (see Figure s10), we found that the core-shell particles can absorb more light at 15 °C than at 40 °C, thereby signifying that the core-shell particles can produce more active OH radicals at the nanocube at the lower temperature.[47,48] The quantitative details of these intricate and very local phenomena are currently difficult to explore and should be interesting for future work.

## Conclusions

In conclusion, we introduce a novel method to synthesize hybrid core-shell microgels consisting of Cu$_2$O nanocubes as the core and thermosensitive PNIPAM as the shell. The core-shell nanoreactors present much higher colloidal stability than pure Cu$_2$O nanocubes in water solution. In addition, the PNIPAM shell can effectively protect the Cu$_2$O nanocubes from oxidation for months. The Cu$_2$O@PNIPAM core-shell nanoreactors show significant enhancement for the photo decomposition of methyl orange under visible light: the reaction rate of the core-shell nanoreactors is 450 times of that of pure Cu$_2$O nanocubes at 15°C. Moreover, temperature can be used as a trigger to control the photocatalytic activity of the Cu$_2$O@PNIPAM core-shell microgels as expected from theory. The present work proves that modification of Cu$_2$O nanocubes with PNIPAM shell will have a great potential for the applications of Cu$_2$O nanoparticles, which is essential to understand the effect of PNIPAM shell on properties of metal or metal oxide nanomaterials.

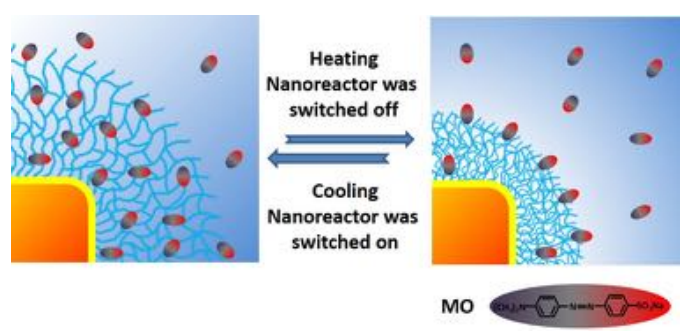

**Figure 6.** Illustration of the catalytic process. At low temperature the network is fully swollen by water and the hydrophilic dye molecules will be enriched within the network. As a consequence of this, the reaction rate for photocatalysis will be increased. In the shrunken state, the gel is in a hydrophobic state that suggests a local repartitioning of MO within the PNIPAM shell: the MO would be enriched in the hydrogel/solvent interface leading to a reduced concentration adjacent to the nanocube. This MO repartitioning will contribute to the diminution of the reaction rate for photocatalysis above the LCST.

## Acknowledgements

S.A.-U. acknowledges the Beijing Municipal Centre for Soft Matter Engineering for funding, as well as the Alexander von Humboldt Foundation (AvH) for an AvH Postdoctoral Research Fellowship. J.D. is grateful for support from the AvH and the ERC (European Research Council) Consolidator Grant with project number 646659 − NANOREACTOR. H. J. gratefully acknowledges the financial support of CSC scholarship. We also thank Dr. Thomas Dittrich (Institute of Heterogeneous Materials, Helmholtz-Center Berlin for materials and energy) for his kind help.

# Supporting information

# Thermosensitive $Cu_2O$-PNIPAM core-shell nanoreactors with tunable photocatalytic activity

He Jia,[1] Rafael Roa,[1] Stefano Angioletti-Uberti,[2,1] Katja Henzler,[3] Andreas Ott,[1] Xianzhong Lin,[4] Jannik Möser,[5] Zdravko Kochovski,[1] Alexander Schnegg,[5] Joachim Dzubiella,[1,6] Matthias Ballauff,[1,6] Yan Lu[1]*

[1]Soft Matter and Functional Materials, Helmholtz-Zentrum Berlin für Materialien und Energie, Hahn-Meitner-Platz 1, Berlin (Germany), [2]International Research Centre for Soft Matter, Beijing University of Chemical Technology, Beisanhuan East Road 19, 100099 Beijing (PR China), [3]Paul Scherrer Institut, 5232 Villigen PSI (Switzerland), [4]Heterogeneous Materialien und Energie, Helmholtz-Zentrum Berlin für Materialien und Energie, Hahn-Meitner-Platz 1, Berlin (Germany), [5]Institute for Nanospectroscopy Helmholtz-Zentrum Berlin für Materialien und Energie, Kekulestr 5, Berlin (Germany), [6]Institut für Physik, Humboldt-Universität zu Berlin, Newtonstr. 15, Berlin (Germany).

*Correspondence and requests for materials should be addressed to Y. L. (email: yan.lu@helmholtz-berlin.de)

## Theory for surface versus diffusion controlled reactions

We have stated in the main manuscript that the reaction is surface controlled. To reach to such conclusion we consider that the total reaction time, $k^{-1}$, is the sum of the time for the reactant methyl orange (MO) to diffuse to the Cu$_2$O nanocubes, $k_D^{-1}$, and the time to react once it is in the surface proximity, $k_S^{-1}$, i.e.,

$$k^{-1} = k_D^{-1} + k_S^{-1} \tag{s1}$$

The time to diffuse to the Cu$_2$O nanocubes is the sum of the time to diffuse from the bulk to the nanoreactor shell, $k_{D0}^{-1}$, and the time to diffuse from the nanoreactor shell to the nanocube, $k_{Dg}^{-1}$, i.e. $k_D^{-1} = k_{D0}^{-1} + k_{Dg}^{-1}$. As we find that MO has a two-orders of magnitude higher concentration (see below) in the hydrogel shell than in bulk, according to our theory for nanoreactors [R1], the rate limiting step in the diffusive approach (i.e. the slowest time) is the mean time of MO to reach the nanoreactor shell. Thus, the diffusion-controlled rate, $k_D$, is simply given by the Smoluchowski equation

$$k_D \approx k_{D0} = 4\pi D_0 R_g c_0 \tag{s2}$$

while we express the surface reaction, $k_S$, as

$$k_S = K_{vol} c_g \tag{s3}$$

where $D_0$ is the MO bulk diffusion coefficient, $R_g$ is the radius of the core-shell nanoreactor, $c_0$ is the bulk MO concentration, $c_g$ is the MO concentration in the hydrogel adjacent to the nanocube surface, and $K_{vol} = k_{vol}\Delta V$, being $k_{vol}$ the fraction per unit time of the MO molecules arriving to the nanocubes that are allowed to react, and $\Delta V$ the volume of the shell next to the nanocubes where effectively the chemical reaction is happening. As a consequence, the surface reaction is directly proportional to the number of reactants within the reactive volume. This also holds naturally for the number of hydroxyl radicals in the reactive volume as well. In the main manuscript we explain that the changes in the reaction rate can be rationalized qualitatively by the changes in the production of OH radicals and by a local re-partitioning of the reactants. Hence, these two effects are included in Eq. (s3) through the local number of reactants and radicals in the reactive volume.

We estimate the diffusion time of the MO reactants to the nanoreactor and we find it to be much faster than the measured total reaction time, which means that the reaction is surface controlled (see Table s1). Note that the diffusion-controlled rate is many orders of magnitudes faster and details of the assumptions leading to eq. (s2) (e.g., the exact value of MO concentration in the hydrogel or which exact shell radius really to take) are negligible.

**Table s1.** $Cu_2O$@PNIPAM core-shell nanoreactors. Total measured reaction rate $k$, theoretical diffusion rate $k_D$ from eq. (s2) and the finally calculated surface reaction rates $k_S$ from using eq. (s1) at different temperatures. The total reaction rates shown here (in units of $s^{-1}$) have been obtained by multiplying the measured ones (in units of $L\ min^{-1}\ m^{-2}$) by the surface of the reacting nanocubes and dividing them by the suspension volume.

| $T$ (°C) | $k$ ($10^{-4}\ s^{-1}$) | $k_D$ ($10^7\ s^{-1}$) | $k_S$ ($10^{-4}\ s^{-1}$) |
|---|---|---|---|
| 15 | 2.4 | 9.5 | 2.4 |
| 40 | 0.19 | 1.1 | 0.19 |

## Average concentration of MO from adsorption data in Fig. s10

The average MO concentration in the PNIPAM shell can be obtained from the adsorption data in Fig. s10 as

$$c_g = c_0 \frac{1}{\phi} \frac{m}{m_0} \tag{s4}$$

where $\phi$ is the nanoreactor volume fraction, and $m$ and $m_0$ are the MO mass adsorbed by the nanoreactors and the total MO mass added to the suspension, respectively (see Fig. s10). The numerical values for our system are shown in Table s2. First, we observe that MO strongly absorbs to the nanoreactors where a two-orders of magnitude higher concentration (on average) than in bulk is found. Thus, nanoreactors feature a large binding affinity of MO in both hydrophilic and hydrophobic cases. Although the amount of MO adsorbed by the nanoreactors is roughly three times larger at 15ºC than at 40ºC, the average MO concentration in the hydrogel is 1.75 times larger at 40ºC as the volume of the PNIPAM shell is approximately fives times smaller. While a larger $c_g$ would give raise to a larger reaction rate, literature suggests that there is a non-trivial local re-partitioning of MO within the PNIPAM shell at 40ºC [R2], where the MO would be mostly situated in a thin shell at the outer surface (the hydrogel/water interface) of the core-shell nanoreactors, leading to a reduced concentration close to the nanoparticle and accordingly a diminished rate.

**Table s2.** Cu$_2$O@PNIPAM core-shell nanoreactors. Volume fraction of nanoreactors, mass of MO adsorbed by the nanoreactors, and ratio between the MO concentrations in the gel and in the bulk at different temperatures.

| $T$ (ºC) | $\phi$ | $m/m_0$ | $c_g/c_0$ |
|---|---|---|---|
| 15 | 4×10$^{-4}$ | 0.075 | 204 |
| 40 | 8×10$^{-4}$ | 0.027 | 358 |

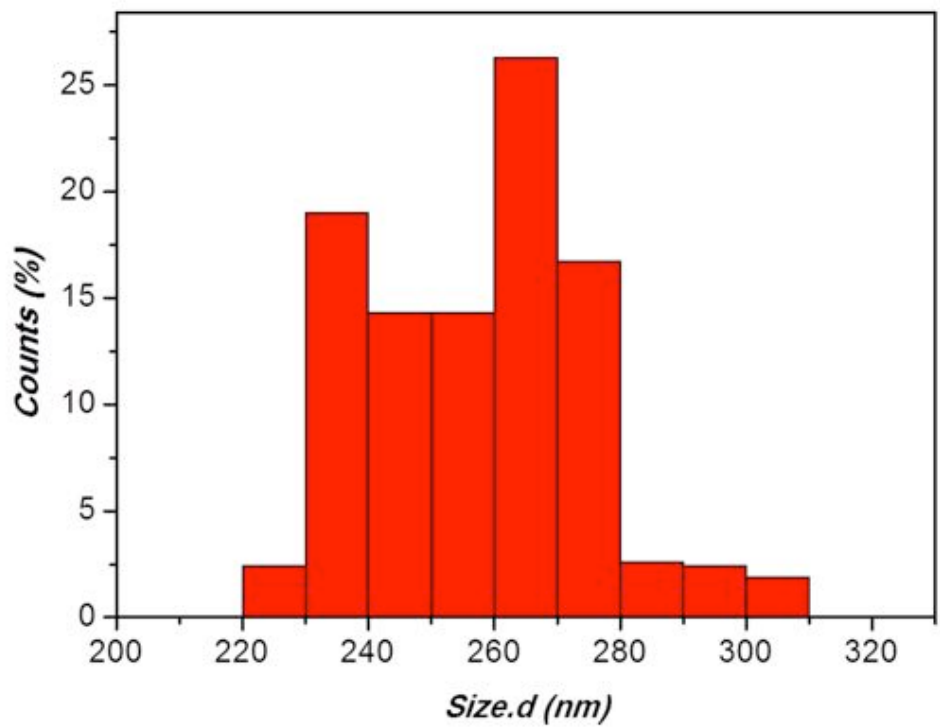

**Figure s1.** The size distribution of pure $Cu_2O$ nanocubes.

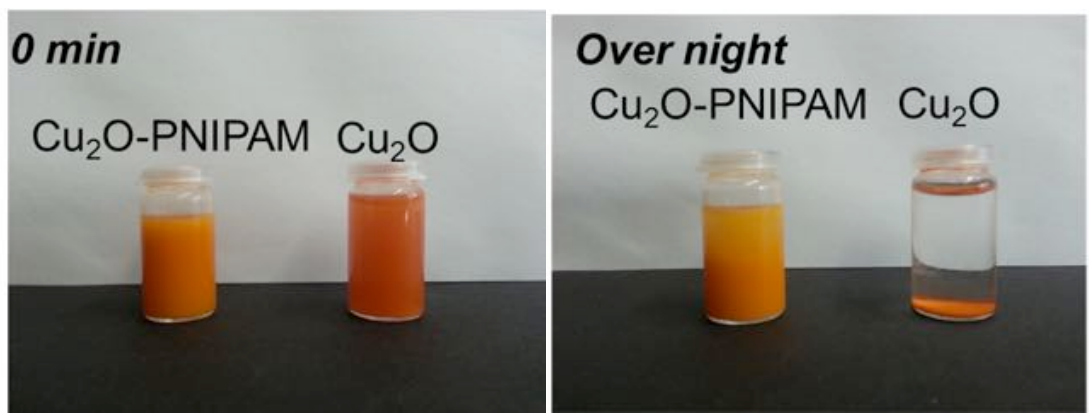

**Figure s2.** The photographs of solution of $Cu_2O$@PNIPAM core-shell nanoparticles (left) and $Cu_2O$ nanocubes (right) after storing for different times.

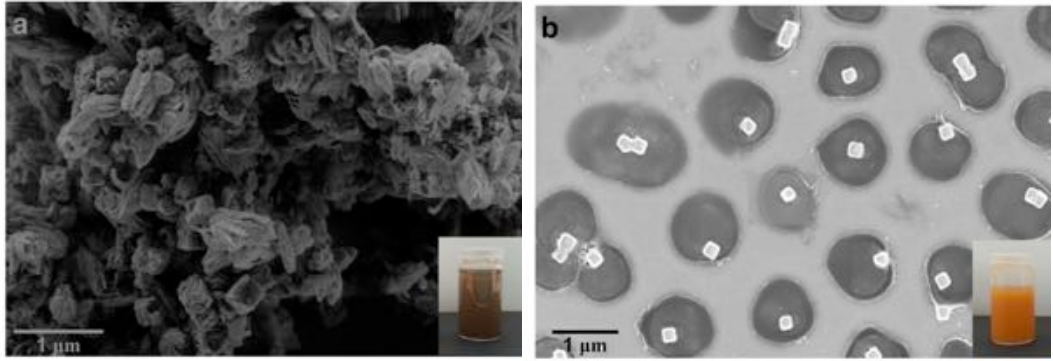

**Figure s3.** SEM images of (a) $Cu_2O$ nanocubes without coating PNIPAM: kept in water at room temperature for 10 days, (b) $Cu_2O$@PNIPAM core-shell nanoparticles: kept in water at room temperature for 10 days.

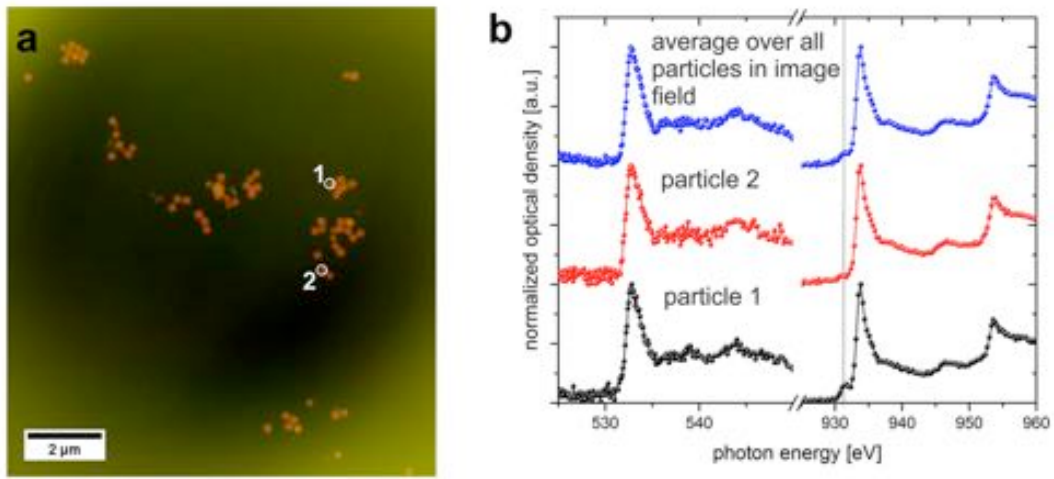

**Figure s4.** (a) TXM micrograph of bare $Cu_2O$ nanocubes at two different photon energies: The red channel depicts the nanocubes with $Cu_2O$, and the green channel refers to CuO. (b) NEXAFS-spectra of the average signals over all $Cu_2O$ nanocubes in field of view (blue line) and the marked particles in the inset of micrograph on the left hand side at the O-$K$-edge and the Cu-$L_{2,3}$-edge (red and black lines).

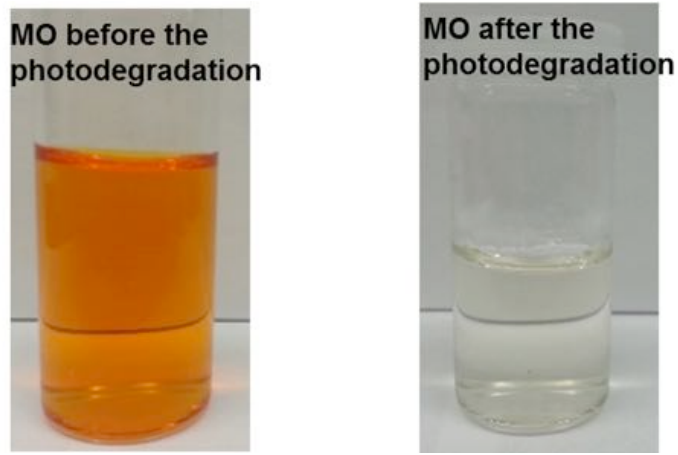

**Figure s5.** The color change of MO before and after photodegradation using Cu2O@PNIPAM core-shell nanocubes as the photocatalyst.

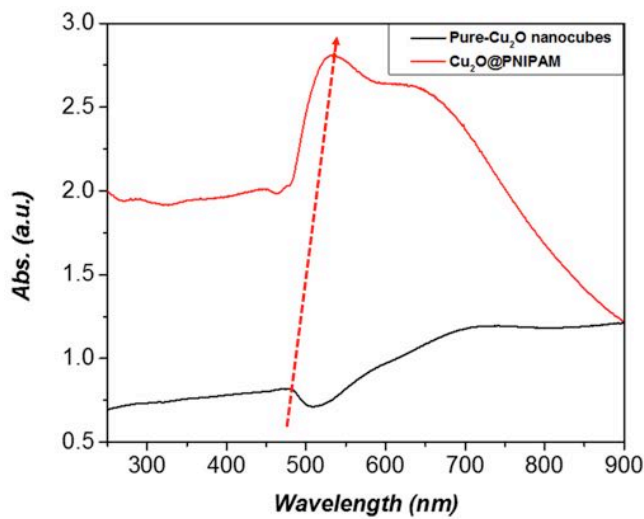

**Figure s6**. UV-vis spectra of $Cu_2O$ nanocubes (black) and $Cu_2O$@PNIPAM (red) at room temperature with the solid content of 0.21mg/mL.

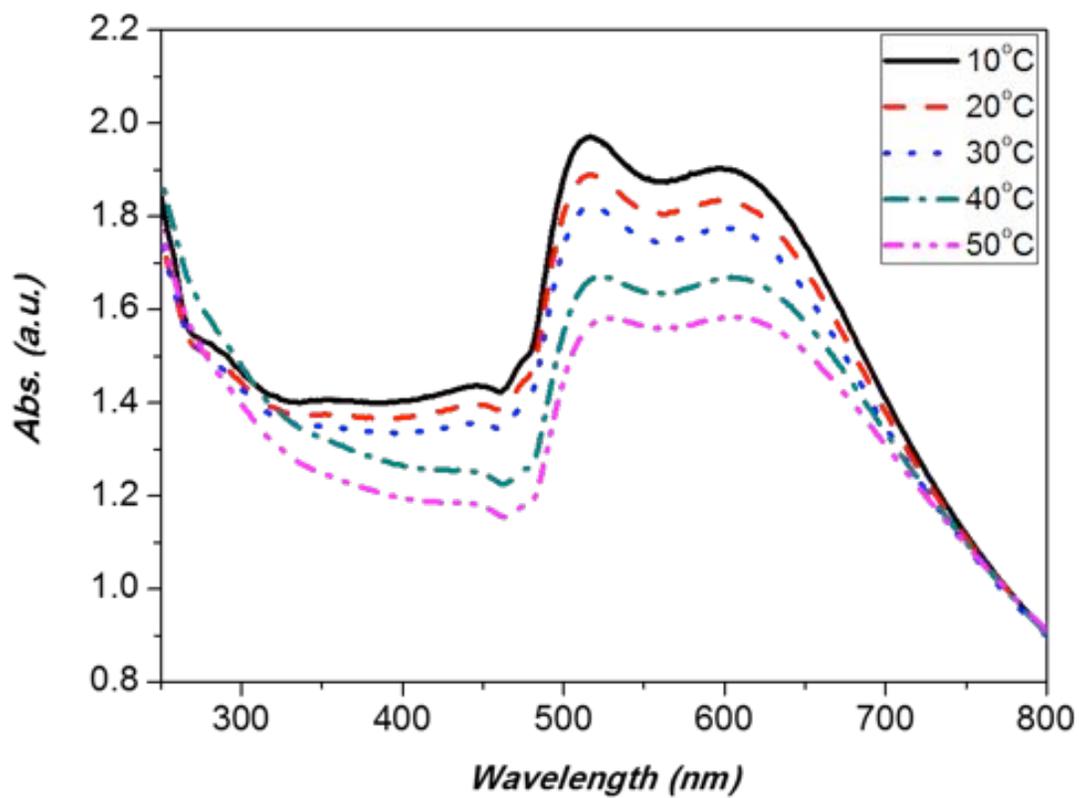

**Figure s7.** UV-vis spectra of $Cu_2O$@PNIPAM core-shell nanoparticles with changing of the temperature.

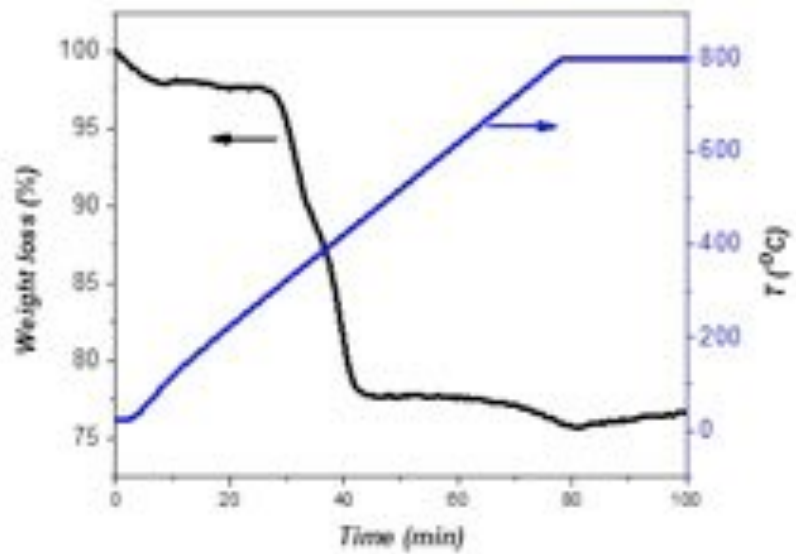

**Figure s8.** TGA spectra of $Cu_2O$@PNIPAM core-shell nanoparticles.

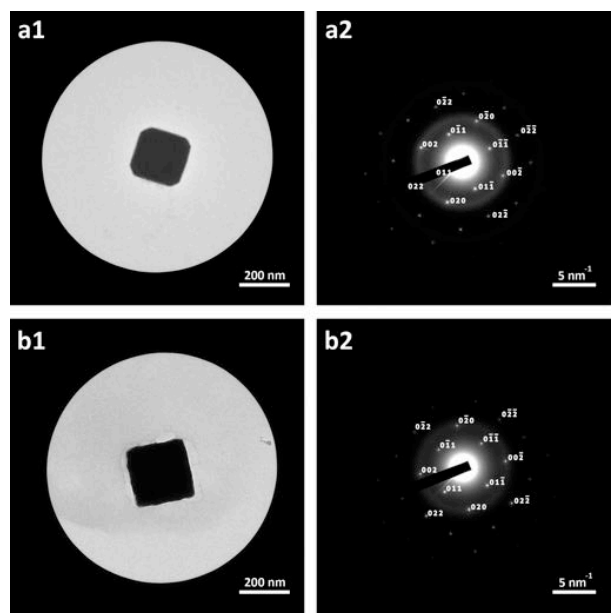

**Figure s9.** TEM images and their corresponding selected area electron diffraction (SAED) patterns of fresh prepared bare $Cu_2O$ nanocubes (a1 and a2), and $Cu_2O$@PNIPAM core-shell nanoparticles (b1 and b2).

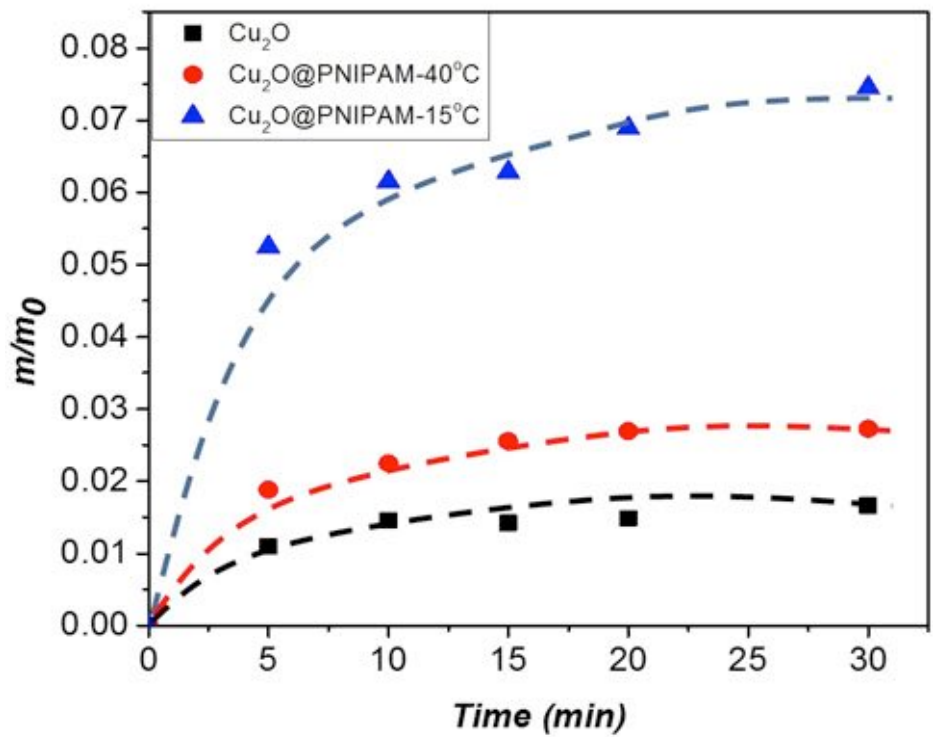

**Figure s10.** The Methyl Orange (MO) adsorption curve of pure $Cu_2O$ nanocubes at 15°C (squares), $Cu_2O$@PNIPAM core-shell nanoparticles at 40°C (circles) and at 15°C (triangles), respectively.